\documentclass[onecolumn,aps,prd,showpacs,eqsecnum]{revtex4}
\usepackage{amsmath,amsfonts,latexsym,epsfig}
\begin{document}
\title{Rotating black hole orbit functionals in the frequency domain}
\author{Steve Drasco}\email{sd68@cornell.edu}
\affiliation{Laboratory for Elementary Particle Physics, Cornell
University, Ithaca, NY 14853}
\author{Scott A.\ Hughes}\email{sahughes@mit.edu}
\affiliation{Department of Physics, MIT, 77 Massachusetts Ave.,
Cambridge, MA 02139}
\begin{abstract}
In many astrophysical problems, it is important to understand the
behavior of functions that come from rotating (Kerr) black hole
orbits.  It can be particularly useful to work with the frequency
domain representation of those functions, in order to bring out their
harmonic dependence upon the fundamental orbital frequencies of Kerr
black holes. Although, as has recently been shown by W.\ Schmidt, such
a frequency domain representation must exist, the coupled nature of a
black hole orbit's $r$ and $\theta$ motions makes it difficult to
construct such a representation in practice.  Combining Schmidt's
description with a clever choice of timelike coordinate suggested by
Y.\ Mino, we have developed a simple procedure that sidesteps this
difficulty.  One first Fourier expands all quantities using Mino's
time parameter $\lambda$.  In particular, the observer's time $t$ is
decomposed with $\lambda$.  The frequency domain description is then
built from the $\lambda$-Fourier expansion and the expansion of $t$.
We have found this procedure to be quite simple to implement, and to
be applicable to a wide class of functionals.  We test the procedure
using a simple test function, and then apply it to a particularly
interesting case, the Weyl curvature scalar $\psi_4$ used in black
hole perturbation theory.
\end{abstract}
\maketitle

\section{Introduction}
\label{sec:intro}

The black holes which appear to exist in a wide range of masses
throughout the universe (see, e.g.\ Refs.\
\cite{bailyn98,begelman2003,korm_geb2001,ferr_2002,cm1999,gebhardt2000,
vdm2001,cp2002})
are most likely described by the Kerr solution of general relativity.
The charged generalization is unlikely to be interesting, as
macroscopic charged objects should be rapidly neutralized by
astrophysical plasma.  The Schwarzschild limit is an unrealistic
idealization given how unlikely it is for an astrophysical macroscopic
object to have precisely zero spin.  This motivates a need
to thoroughly understand phenomena in the vicinity of Kerr black
holes.  Such an understanding becomes quite important as studies
probe ever more deeply into black holes' strong fields.

Of particular interest to many applications is an understanding of
Kerr black hole orbits.  In the language of general relativity,
``orbits'' are bound, stable geodesic trajectories.  It is a
relatively simple matter to write down the equations governing these
orbits and to integrate in the time domain to find the detailed
trajectory that a body will follow.

These orbits have a rich phenomenology, owing to the complicated shape
of the hole's gravitational ``potential''.  At largish radii ($r
\gtrsim 20$ times the radius of the hole), a generic orbit is not too
different from the ellipses of Newtonian theory.  However, the plane
in which this ellipse lies precesses (due largely to the spin of the
black hole and the oblateness of the hole's geometry), and the ellipse
precesses within that precessing plane.  We can identify two
fundamental orbital frequencies: a frequency $\Omega_r$ characterizing
the radial motion (from periapsis to apoapsis and back), and a
frequency $\Omega_\theta$ characterizing the latitudinal motion.  A
third frequency of somewhat different nature describes the average
secular accumulation of the angle about the hole's symmetry axis, and
is denoted by $\Omega_\phi$.  The various precessions of the orbit are
due to mismatches between these frequencies: the orbital plane
precesses at $\Omega_\phi - \Omega_\theta$; the orbital ellipse
precesses at $\Omega_\phi - \Omega_r$.  Closed form expressions for
all three of these frequencies have recently been worked out by W.\
Schmidt {\cite{ws}}.  In the deep strong field of the hole, the
frequencies become so different that the qualitative picture given
above --- a precessing ellipse on a precessing plane --- ceases to be
useful.  The orbits just become complicated and messy.

Despite this complicated nature, a wide class of functions of black
hole orbits are completely described by the frequencies $\Omega_r$ and
$\Omega_\theta$.  Any function of the form $f[r(t),\theta(t)]$ (a
common functional form for black hole orbits, since the metric is
independent of both $t$ and $\phi$) can be expanded as
\begin{equation}
f[r(t),\theta(t)] = \sum_{kn} f_{kn}
e^{-ik\Omega_\theta t}e^{-in\Omega_r t}\;.
\label{eq:f_expand_t}
\end{equation}
Unless otherwise noted, the index of all sums runs from $-\infty$ to
$\infty$.  The fact that such expansions exist is very useful, since
it suggests we can Fourier analyze a wide class of interesting orbit
functionals to understand their harmonic dependence upon the orbital
frequencies.

Some functions have a more complicated form depending on all four
components of the orbital worldline, $z^\alpha = (t,r,\theta,\phi)$.
A similar, but slightly modified, expansion can be written down which
handles functions of this sort.  Such an expansion is needed, for
example, to give the harmonic decomposition of an orbiting body's
stress-energy tensor, used in frequency domain perturbation theory of
Kerr black holes {\cite{teuk73}}.  One could also imagine using this
harmonic expansion to describe the emission spectrum of hot material
accreting onto a black hole.  This could facilitate identifying
features that are imprinted upon a black hole's x-ray spectrum.

Actually computing the expansion coefficients $f_{kn}$ turns out to be
somewhat difficult.  This is fundamentally because the $r$ and
$\theta$ motions of a black hole orbit are coupled, and as a result
are not periodic in coordinate time $t$ (or proper time $\tau$).  This
difficulty can be fixed by working with a time variable $\lambda$,
recently suggested by Y.\ Mino {\cite{mino}}, which decouples the $r$
and $\theta$ motions.  With respect to $\lambda$, the $r$ and $\theta$
motions are truly periodic.  In contrast to the time domain expansion
(\ref{eq:f_expand_t}), a similar expansion using $\lambda$ is
straightforward to compute.

The clocks of distant observers tick at evenly spaced intervals of
$t$, not $\lambda$.  For the purpose of describing quantities that
could be measured by such observers, the $t$ expansion is more useful
than the $\lambda$ expansion.  Fortunately, it is straightforward to
convert.  That is the subject of this paper.  The key observation is
that observer time $t$ contains oscillatory elements that are periodic
with respect to Mino's time $\lambda$.  Thus, $t$ itself can be
expanded in a Fourier series of $\lambda$-frequency harmonics.  

The remainder of this paper describes our prescription.  In Sec.\
{\ref{sec:BL}}, we briefly discuss the $t$-domain description of 
the orbits.  We then show how Mino's time $\lambda$ fixes many of the 
difficulties associated with these orbits in Sec.\ {\ref{sec:Mino}}.  
In Sec.\ {\ref{sec:convert}}, we show how to use a $\lambda$ expansion 
to compute the the $t$ expansion coefficients $f_{kn}$.  In 
Sec.\ {\ref{sec:example}}, we apply this technique first to a relatively 
simple function of black hole orbits, and then to the Weyl curvature 
scalar $\psi_4$, demonstrating that everything works quite robustly.  
Appendix {\ref{app:practical}} discusses some important details related 
to implementation of these techniques.

\section{Orbits in Boyer-Lindquist time}
\label{sec:BL}

The geodesic equations that govern Kerr black hole orbits are usually
presented in the following ``classic'' form {\cite{mtw}:
\begin{eqnarray}
\rho^4\left(\frac{dr}{d\tau}\right)^2 &=& \left[E(r^2+a^2)
- a L_z\right]^2- \Delta\left[r^2 + (L_z - a E)^2 +
Q\right]
\equiv R(r)\;,
\label{eq:rdot}\\
\rho^4\left(\frac{d\theta}{d\tau}\right)^2 &=& Q - \cot^2\theta L_z^2
-a^2\cos^2\theta(1 - E^2)
\equiv\Theta(\theta)\;,
\label{eq:thetadot}\\
\rho^2\left(\frac{d\phi}{d\tau}\right) &=&
\csc^2\theta L_z + aE\left(\frac{r^2+a^2}{\Delta} - 1\right) -
\frac{a^2L_z}{\Delta}
\equiv\Phi(r,\theta)\;,
\label{eq:phidot}\\
\rho^2\left(\frac{dt}{d\tau}\right) &=&
E\left[\frac{(r^2+a^2)^2}{\Delta} - a^2\sin^2\theta\right] +
aL_z\left(1 - \frac{r^2+a^2}{\Delta}\right)
\equiv T(r,\theta)\;.
\label{eq:tdot}
\end{eqnarray}
Up to initial conditions, orbits are specified by the quantities $E$,
$L_z$, and $Q$ (``energy'', ``$z$-component of angular momentum'', and
``Carter constant''); these quantities are conserved along any orbit
of the family.  For notational simplicity, we have put $\rho^2 = r^2 +
a^2\cos^2\theta$ and $\Delta = r^2 - 2 M r + a^2$.  Note that Eqs.\
(\ref{eq:rdot}) and (\ref{eq:thetadot}) have been divided by $\mu^2$,
and Eqs.\ (\ref{eq:phidot}) and (\ref{eq:tdot}) by $\mu$ (where $\mu$
is the mass of a small body in an orbit); $E$, $L_z$, and $Q$ are thus
the specific energy, angular momentum and Carter constant.  The
parameter $\tau$ is proper time measured along the orbit; $t$ is
Boyer-Lindquist coordinate time.  We choose $0 \le a \le M$; prograde
and retrograde orbits are distinguished by an orbital inclination
angle rather than the sign of the hole's spin.

By picking initial conditions and physically reasonable values of the
constants $E$, $L_z$, and $Q$, one can integrate these equations to
obtain a worldline parameterized by proper time $\tau$ along the
orbit.  Schmidt {\cite{ws}} has derived formulae for these constants
as functions of an orbit's semi-latus rectum $p$, eccentricity $e$,
and an inclination angle $\iota$; further discussion of these
parameters is given in Appendix {\ref{app:practical}}.  Schmidt's
formulae do not work well for circular orbits ($e = 0$).  Formulae
which apply to that case were originally worked out by Shakura
{\cite{ns}}; we use a parameterization which was originally derived by
Williams {\cite{rw}}, and then re-derived by Hughes {\cite{h2001}}.

For the purpose of understanding quantities which could be measured by
distant observers, proper time is not a particularly good choice of
parameterization for the orbit --- it is connected to the orbit
itself, and so contains components which oscillate with respect to the
clocks of distant observers.  Since the Boyer-Lindquist time
coordinate $t$ reduces at large radius to time as measured by distant
observers, one should parameterize with $t$ rather than $\tau$.  It is
trivial to convert: just divide the geodesic equations in $\tau$ by
$dt/d\tau$ to obtain equations in $t$:
\begin{equation}
\frac{dr}{dt} = \frac{dr}{d\tau}\left(\frac{dt}{d\tau}\right)^{-1}\;,
\end{equation}
and likewise for $d\theta/dt$ and $d\phi/dt$.  Then, pick initial
conditions and an allowed set of orbital constants ($E,L_z,Q$), and
integrate to find ${\bf z}(t) = [r(t),\theta(t),\phi(t)]$.

Using elegant Hamilton-Jacobi techniques, W.\ Schmidt {\cite{ws}} has
recently shown that bound orbits satisfying these equations are
characterized by multiply-periodic motion in $r$, $\theta$, and $\phi$.
These motions are given by three fundamental frequencies, $\Omega_r$,
$\Omega_\theta$, and $\Omega_\phi$.
In fact, the frequency $\Omega_\phi$ can be considered
less fundamental than $\Omega_r$ and $\Omega_\theta$.  This is because
the $\phi$ orbital motion corresponds (in the language of Goldstein
{\cite{goldstein}}) to a {\it rotation}-type periodic motion, rather
than an oscillatory or {\it libration}-type periodicity.  The
frequency $\Omega_\phi$ is the average rate at which $\phi$
accumulates over an orbit.  Because $d\phi/dt$ depends only on $r$ and
$\theta$, deviations from that average accumulation are oscillations
at the $r$ and $\theta$ frequencies:
\begin{equation}
\phi(t) = \Omega_\phi t + \sum_{kn}\varphi_{kn} e^{-ik\Omega_\theta t}
e^{-in\Omega_r t}\;.
\end{equation}
Physically, one can imagine analyzing black hole orbits in a frame
that co-rotates at the frequency $\Omega_\phi$. In that corotating
frame, the rotation-type periodicity at $\Omega_\phi$ is removed, and
only the libration-type oscillations at harmonics of $\Omega_r$ and
$\Omega_\theta$ remain (see also discussion in Ref.\
{\cite{goldstein}}, pp.\ 466 -- 467).

By this logic, many functions $f[{\bf z}(t)]$ can be reduced to
functions of $r$ and $\theta$ only.  It is then possible to expand in
a Fourier series as
\begin{equation}
f[r(t),\theta(t)] = \sum_{kn}f_{kn} e^{-i\left(k\Omega_\theta + 
n\Omega_r\right)t}\;.
\label{eq:expand}
\end{equation}
Unfortunately, the functions $r(t)$ and $\theta(t)$ are in general not
periodic (although they are in the Newtonian limit where all the
orbital frequencies are identical).  This not-quite-periodic character
is fundamentally due to the coupling the $r$ and $\theta$ motions in
Eqs.~(\ref{eq:rdot}) and (\ref{eq:thetadot}): the functions $(\rho^2
dt/d\tau)^{-2}R$ and $(\rho^2 dt/d\tau)^{-2}\Theta$ each depend
explicitly on both $r$ and $\theta$.  (Note that this coupling remains
if we use proper time along the orbit $\tau$ as our parameterization.)
The non-separated nature of the $r$ and $\theta$ motions makes it
difficult to compute the coefficients $f_{kn}$ appearing in Eq.\
(\ref{eq:expand}).  If the motions separated, one could define angle
variables $w^r\equiv\Omega_r t$ and $w^\theta\equiv\Omega_\theta t$,
such that $r$ would be a function only of $w^r$ and $\theta$ a
function only of $w^\theta$ {\cite{goldstein,schwarzschild}}.
Computing the coefficients $f_{kn}$ would then be straightforward
(see, e.g., Ref.\ {\cite{goldstein}}, p.\ 466).  Since the motions do
not in fact separate, the angles $w^r$ and $w^\theta$ are not well
defined.  An alternative scheme to compute the Fourier series
coefficients appears necessary.

\section{Orbits in Mino time}
\label{sec:Mino}

In a recent paper, Y.\ Mino {\cite{mino}} introduced a new
parameterization of Kerr geodesic motion which separates the $r$ and
$\theta$ motion.  In terms of what we shall call ``Mino time''
$\lambda$, the geodesic equations become
\begin{eqnarray}
\left(\frac{dr}{d\lambda}\right)^2 &=& R(r)\;,
\label{eq:rlamb}\\
\left(\frac{d\theta}{d\lambda}\right)^2 &=& \Theta(\theta)\;,
\label{eq:thetalamb}\\
\frac{d\phi}{d\lambda} &=& \Phi(r,\theta)\;,
\label{eq:dphidlamb}\\
\frac{dt}{d\lambda} &=& T(r,\theta)\;,
\label{eq:dtdlamb}
\end{eqnarray}
where $R(r)$, $\Theta(\theta)$, $\Phi(r,\theta)$, and $T(r,\theta)$
are defined in Eqs.\ (\ref{eq:rdot})--(\ref{eq:tdot}).  The $r$ and
$\theta$ motions are now strictly periodic functions:
\begin{eqnarray}
r(\lambda) &=& r(\lambda + n\Lambda_r)\;,
\nonumber\\
\theta(\lambda) &=& \theta(\lambda + n\Lambda_\theta)\;,
\label{eq:lambdaperiodic}
\end{eqnarray}
where $n$ is any integer and the periods are given by
\begin{eqnarray}
\Lambda_r &=& 2\int_{r_{\rm peri}}^{r_{\rm ap}}\frac{dr}{R(r)^{1/2}}\;,
\label{eq:LambdaR}\\
\Lambda_\theta &=& 4\int_{\theta_{\rm min}}^{\pi/2}
\frac{d\theta}{\Theta(\theta)^{1/2}}\;.
\label{eq:LambdaTheta}
\end{eqnarray}
The radial motion is taken to range between periapsis, $r_{\rm peri}$,
and apoapsis, $r_{\rm ap}$; the $\theta$ motion ranges from a minimum
$\theta_{\rm min}$ to a maximum $\pi - \theta_{\rm min}$.  (With a
particular reparameterization, we can write the $\Lambda_r$ integral
in such a way that it behaves well as we approach the limit of
circular orbits, $r_{\rm peri} \to r_{\rm ap}$.  Likewise it is simple
to reparameterize such that $\Lambda_\theta$ is well behaved in the
equatorial orbit limit, $\theta_{\rm min}\to\pi/2$.  See Appendix
{\ref{app:practical}}.)

For what follows, it will be useful to define the following
frequencies conjugate to $\lambda$:
\begin{equation} \label{eq:freqdef} 
\Upsilon_{r,\theta} = 2\pi/\Lambda_{r,\theta},
\end{equation} 
as well as the angle variables
\begin{equation}
w^{r,\theta} = \Upsilon_{r,\theta}\lambda\;.
\label{eq:angledef}
\end{equation}
These angles allow us to take advantage of the separated nature of $r$
and $\theta$ motion in Mino time: we treat $r$ as a function only of
$w^r$, $\theta$ as a function only of $w^\theta$, and we treat $w^r$
and $w^\theta$ as independent parameters.  This allows us to Fourier
decompose any function of the orbital worldline using standard
action-angle variable techniques {\cite{goldstein}}.

Before moving on, we should analyze the remaining coordinate motions
of black hole orbits --- the observer (Boyer-Lindquist) time $t$ and
the azimuthal angle $\phi$.  Both of these motions consist of a
component that accumulates secularly as a function of $\lambda$,
superposed on components which oscillate at $\Upsilon_r$ and
$\Upsilon_\theta$.  Let us analyze the oscillations first.  From the
geodesic equations (\ref{eq:dphidlamb}) and (\ref{eq:dtdlamb}), we
know that $dt/d\lambda$ and $d\phi/d\lambda$ are functions only of $r$
and $\theta$.  This means that they can be expanded in a Fourier
series:
\begin{eqnarray}
\frac{dt}{d\lambda} \equiv T(r,\theta) &=& \sum_{kn} T_{kn}
e^{-i\left(k\Upsilon_\theta + n\Upsilon_r\right)\lambda}\;,
\label{eq:dtdlamb_expand}\\
\frac{d\phi}{d\lambda} \equiv \Phi(r,\theta) &=& \sum_{kn}\Phi_{kn}
e^{-i\left(k\Upsilon_\theta + n\Upsilon_r\right)\lambda}\;,
\label{eq:dphidlamb_expand}
\end{eqnarray}
with the expansion coefficients given by
\begin{eqnarray}
T_{kn} &=& \frac{1}{(2\pi)^2}\int_0^{2\pi} dw^r\int_0^{2\pi}
dw^\theta\, T\left[r(w^r),\theta(w^\theta)\right]e^{i\left(kw^\theta +
nw^r\right)}\;,
\label{eq:t_coeff}\\
\Phi_{kn} &=& \frac{1}{(2\pi)^2}\int_0^{2\pi} dw^r\int_0^{2\pi}
dw^\theta\, \Phi\left[r(w^r),\theta(w^\theta)\right]e^{i\left(kw^\theta +
nw^r\right)}\;.
\label{eq:phi_coeff}
\end{eqnarray}
In these equations and in what follows, $r(w^r) \equiv r(\lambda =
w^r/\Upsilon_r)$ and $\theta(w^\theta) \equiv \theta(\lambda =
w^\theta/\Upsilon_\theta)$.

Because the functions $T(r,\theta)$ and $\Phi(r,\theta)$ are real,
we have the following relations:
\begin{eqnarray}
T_{-k,-n} &=& {\bar T}_{kn}\;,
\label{eq:Tconjugate}\\
\Phi_{-k,-n} &=& {\bar \Phi}_{kn}\;,
\label{eq:Phiconjugate}
\end{eqnarray}
where the overbar denotes complex conjugation.  The matrices $T_{kn}$
and $\Phi_{kn}$ have another interesting property: $T_{k0}$ and
$T_{0n}$ are non-zero, but $T_{kn} = 0$ if $k \ne 0$ and $n \ne 0$
(and likewise for $\Phi_{kn}$).  This lack of ``crosstalk'' between
the $\theta$ and $r$ harmonics is because $T(r,\theta)$ and
$\Phi(r,\theta)$ have the form $f(r) + g(\theta)$.  To take advantage 
of this property, we define
\begin{eqnarray}
T^\theta_k &\equiv& T_{k0}\;,\qquad T^r_n\equiv T_{0n}\;;
\label{eq:Tthetar}\\
\Phi^\theta_k &\equiv& \Phi_{k0}\;,\qquad\Phi^r_n\equiv\Phi_{0n}\;.
\label{eq:Phithetar}
\end{eqnarray}
Using the complex conjugate relations and Eqs.\ (\ref{eq:Tthetar}) and
(\ref{eq:Phithetar}), we rewrite the double sums appearing in the
Fourier expansions (\ref{eq:dphidlamb_expand}) and
(\ref{eq:dtdlamb_expand}) as a pair of single sums \cite{sums}:
\begin{eqnarray}
\frac{dt}{d\lambda} &\equiv& T(r,\theta) =
\Gamma +
\sum_{k=1}^\infty\left(T^\theta_k e^{-ik\Upsilon_\theta\lambda} +
\text{c.c.}\right) +
\sum_{n=1}^\infty\left(T^r_n e^{-in\Upsilon_r\lambda} +
\text{c.c.}\right)\;;
\label{eq:dtdlamb_expand2}\\
\frac{d\phi}{d\lambda} &\equiv& \Phi(r,\theta) =
\Upsilon_\phi +
\sum_{k=1}^\infty\left(\Phi^\theta_k e^{-ik\Upsilon_\theta\lambda} +
\text{c.c.}\right) +
\sum_{n=1}^\infty\left(\Phi^r_n e^{-in\Upsilon_r\lambda} +
\text{c.c.}\right)\;.
\label{eq:dphidlamb_expand2}
\end{eqnarray}
The ``c.c.'' means the complex conjugate of the preceding term.  We
have pulled the $k = 0$, $n = 0$ terms out of these sums and defined
\begin{eqnarray}
\Gamma &=& T_{00}\;,
\label{eq:LambdaT}\\
\Upsilon_\phi &=& \Phi_{00}\;.
\label{eq:LambdaPhi}
\end{eqnarray}
These numbers tell us about the secular, average rate at which $\phi$
and $t$ accumulate with respect to $\lambda$.  

Using these results, it is simple to integrate for $\phi(\lambda)$ and
$t(\lambda)$:
\begin{eqnarray}
t(\lambda) &=& \Gamma\lambda + \Delta t(\lambda)\;,\label{eq:tlamb}\\
\phi(\lambda) &=& \Upsilon_\phi\lambda + \Delta\phi(\lambda)\;.
\label{eq:philamb}
\end{eqnarray} 
We have chosen $t(\lambda=0) = 0 = \phi(\lambda=0)$, and defined
\begin{eqnarray} 
\Delta t(\lambda) &=& \sum_{k=1}^\infty \left(
\Delta t^\theta_k e^{-ik\Upsilon_\theta\lambda} + 
\text{c.c.}\right) +
\sum_{n=1}^\infty\left(
\Delta t^r_n e^{-in\Upsilon_r\lambda} + 
\text{c.c.} \right)\;;
\label{eq:deltatlamb}\\
\Delta\phi(\lambda) &=& \sum_{k=1}^\infty \left(
\Delta \phi^\theta_k e^{-ik\Upsilon_\theta\lambda} +
\text{c.c.} \right) +
\sum_{n=1}^\infty\left(
\Delta \phi^r_n e^{-in\Upsilon_r\lambda} +
\text{c.c.}\right)\;.
\label{eq:deltaphilamb}
\end{eqnarray}
We have defined $\Delta t^{r,\theta}_j = i T^{r,\theta}_j/(j
\Upsilon_{r,\theta})$ and $\Delta \phi^{r,\theta}_j =
i\Phi^{r,\theta}_j/(j\Upsilon_{r,\theta})$.  With this definition, we
have separated $t(\lambda)$ and $\phi(\lambda)$ into pieces which
accumulate secularly with $\lambda$ plus pieces $\Delta t(\lambda)$
and $\Delta\phi(\lambda)$ that oscillate at harmonics of
$\Upsilon_\theta$ and $\Upsilon_r$.

Since $\Omega_\phi$ is the average rate at which $\phi$ accumulates as
a function of $t$ and since $\Gamma$ and $\Upsilon_\phi$ are the
average rates at which $t$ and $\phi$ accumulate as a functions of
$\lambda$,
\begin{equation}
\Omega_\phi = \Upsilon_\phi/\Gamma\;.
\label{eq:OmegaPhi}
\end{equation}
The other frequencies are likewise related:
\begin{eqnarray}
\Omega_\theta &=& \Upsilon_\theta/\Gamma\;,
\label{eq:OmegaTheta}\\
\Omega_r &=& \Upsilon_r/\Gamma\;.
\label{eq:OmegaR}
\end{eqnarray}

When performing a harmonic decomposition of any function, we will want
to work in terms of the angles $w^j= \Upsilon_j\lambda$, for $j =
r,\theta,\phi$, and the average accumulated time ${\cal T} =
\Gamma\lambda$.  In terms of these variables,
\begin{eqnarray} 
t({\cal T},w^\theta,w^r) &=& {\cal T} + \Delta t(w^\theta,w^r), \\
\phi(w^\phi,w^\theta,w^r) &=& w^\phi + \Delta\phi(w^\theta,w^r),
\end{eqnarray} 
where
\begin{eqnarray} 
\Delta t(w^\theta,w^r) &=& \sum_{k=1}^\infty \left(
\Delta t^\theta_k e^{-ikw^\theta} + \text{c.c.}\right) +
\sum_{n=1}^\infty
\left(\Delta t^r_n e^{-inw^r} +
\text{c.c.} \right);,
\label{eq:deltat_w}\\
\Delta\phi(w^\theta,w^r) &=& \sum_{k=1}^\infty \left(
\Delta\phi^\theta_k e^{-ikw^\theta} + \text{c.c.}\right) +
\sum_{n=1}^\infty
\left(\Delta\phi^r_n e^{-inw^r} + \text{c.c.} \right)\;.
\label{eq:deltaphi_w}
\end{eqnarray} 

Putting all of this together, the Fourier expansion coefficients
${\tilde f}_{kn}$ of any function of the form
$f[r(\lambda),\theta(\lambda)]$ is
\begin{equation}
{\tilde f}_{kn} = \frac{1}{(2\pi)^2}
\int_0^{2\pi} dw^r~ \int_0^{2\pi} dw^\theta~
f[r(w^r),\theta(w^\theta)]
e^{i\left(kw^\theta + nw^r\right)}\;,
\label{eq:ftilde_def}
\end{equation}

It is useful to note that the worldline $z^\alpha$ can be reorganized
in a similar form, separating the oscillations from the secular
accumulations:
\begin{equation}
z^\alpha(\lambda) = z^\alpha_{\text{sec}}(\lambda) + 
\Delta z^\alpha[r(\lambda),\theta(\lambda)]\;,
\end{equation} 
where $z_{\text{sec}}^\alpha(\lambda) =
(\Gamma\lambda,0,0,\Upsilon_\phi\lambda)$, and where
\begin{equation} 
\Delta z^\alpha[r,\theta] = [\Delta t(r,\theta), r, \theta, \Delta
\phi(r,\theta)]
\end{equation} 
can be expanded using the simple Fourier coefficients described by
Eq.\ (\ref{eq:ftilde_def}) with $f = z^\alpha$.  This leaves the
worldline in the desirable form
\begin{equation} \label{eq:full_z_decomp}
z^\alpha(\lambda) = z^\alpha_{\text{sec}}(\lambda) + \sum_{kn} \Delta
z^\alpha_{kn} e^{-i(k\Upsilon_\theta + n\Upsilon_r)\lambda}\;.
\end{equation} 
By making use of Eq.\ (\ref{eq:full_z_decomp}), even rather
complicated functional forms turn out to have a straightforward
harmonic description.

\section{Converting Fourier expansion coefficients}
\label{sec:convert}

There are two ways of Fourier expanding a function of the form 
$f[r(t),\theta(t)]$ which are essentially equivalent: we can expand in
observer time $t$,
\begin{equation}
f[r(t),\theta(t)] = \sum_{kn} f_{kn} e^{-i\Omega_{kn}t}\;;
\label{eq:f_t_expand}
\end{equation}
or, we can expand in Mino time $\lambda$,
\begin{equation}
f[r(\lambda),\theta(\lambda)] = \sum_{kn} {\tilde f}_{kn}
e^{-i\Upsilon_{kn}\lambda}\;.
\label{eq:f_lamb_expand}
\end{equation}
We have defined
\begin{eqnarray}
\Omega_{kn} &=& k\Omega_\theta + n\Omega_r\;,
\nonumber\\
\Upsilon_{kn} &=& k\Upsilon_\theta + n\Upsilon_r\;.
\label{eq:freq_mkn}
\end{eqnarray}
From the standpoint of measurable physics, the expansion
(\ref{eq:f_t_expand}) is more interesting --- the components $f_{kn}$
tell us about the harmonic structure of $f$ as seen by distant
observers.  However, the expansion (\ref{eq:f_lamb_expand}) is far
more accessible --- using Eq.\ (\ref{eq:ftilde_def}), it is
straightforward to compute the expansion components ${\tilde
f}_{kn}$.  In this section, we show how to convert the accessible
components ${\tilde f}_{kn}$ into the measurable components
$f_{kn}$.

We begin by taking the Fourier transform of $f[r(t),\theta(t)]$.
Using Eq.\ (\ref{eq:f_t_expand}), we have
\begin{eqnarray}
\sum_{kn} f_{kn}\delta(\omega - \Omega_{kn})
&=& \frac{1}{2\pi}\int_{-\infty}^\infty dt\,f[r(t),\theta(t)] e^{i\omega
t}\;,
\nonumber\\
&=& \frac{1}{2\pi}\int_{-\infty}^\infty
d\lambda\,\frac{dt}{d\lambda}\, f[r(\lambda),\theta(\lambda)] e^{i\omega
t(\lambda)}\;.
\label{eq:relate1}
\end{eqnarray}
Our goal is to evaluate the integral on the right-hand side of
(\ref{eq:relate1}) and to find an expression relating $f_{kn}$ to
${\tilde f}_{kn}$.  To do so, we take advantage of the Fourier
expansion for $t(\lambda)$ previously established, (\ref{eq:tlamb}).

We now insert $dt/d\lambda = T(r,\theta)$ and $e^{i\omega t(\lambda)}
= e^{i\omega\Gamma\lambda}\times e^{i\omega\Delta t(\lambda)}$, under
the integral:
\begin{eqnarray}
\sum_{kn} f_{kn}\delta(\omega - \Omega_{kn}) &=&
\frac{1}{2\pi}\int_{-\infty}^\infty d\lambda\,
\left\{T[r(\lambda),\theta(\lambda)]
f[r(\lambda),\theta(\lambda)] e^{i\omega\Delta t(\lambda)}\right\}
e^{i\omega\Gamma\lambda}\;,
\nonumber\\
&\equiv& \frac{1}{2\pi}\int_{-\infty}^\infty d\lambda\,
{\cal F}[r(\lambda),\theta(\lambda),\omega]\,e^{i\omega\Gamma\lambda}\;.
\label{eq:fourier1}
\end{eqnarray}

We next wish to insert the Fourier expansion of ${\cal F}[r(\lambda),
\theta(\lambda),\omega]$ under the integral.  We first write this function
in terms of the angle variables:
\begin{equation}
{\cal F}(w^\theta,w^r,\omega) = T[r(w^r),\theta(w^\theta)]
e^{i\omega\Delta t(w^\theta,w^r)} f[r(w^r),\theta(w^\theta)]\;.
\label{eq:calF_w}
\end{equation}
The expansion of ${\cal F}[r(\lambda),\theta(\lambda),\omega]$ is
\begin{equation}
{\cal F}[r(\lambda),\theta(\lambda),\omega] = \sum_{ab}{\cal F}_{ab}(\omega)
e^{-i\Upsilon_{ab}\lambda}\;,
\label{eq:calF_expand}
\end{equation}
where
\begin{equation}
{\cal F}_{ab}(\omega) = \frac{1}{(2\pi)^2}
\int_0^{2\pi} dw^r~\int_0^{2\pi} dw^\theta~
{\cal F}(w^\theta,w^r,\omega)\,
e^{iaw^\theta}e^{ibw^r}\;.
\label{eq:calF_abc}
\end{equation}
Inserting the expansion (\ref{eq:calF_w}) into Eq.\
(\ref{eq:fourier1}), we find
\begin{eqnarray}
\sum_{kn} f_{kn}\delta(\omega - \Omega_{kn})
&=& \frac{1}{2\pi}\int_{-\infty}^\infty d\lambda\, \sum_{ab}{\cal
F}_{ab}(\omega)\,e^{i(\omega\Gamma - \Upsilon_{ab})\lambda}\;,
\nonumber\\
&=& \sum_{ab}{\cal F}_{ab}(\omega)\,\delta(\omega\Gamma -
\Upsilon_{ab})\;,
\nonumber\\
&=& \Gamma^{-1}\sum_{ab}{\cal F}_{ab}(\omega)\,
\delta(\omega - \Omega_{ab})\;,
\nonumber\\
&=& \Gamma^{-1}\sum_{ab}{\cal F}_{ab}(\Omega_{ab})\,
\delta(\omega - \Omega_{ab})\;.
\label{eq:fourier2}
\end{eqnarray}
Equating the left-hand sides and right-hand sides, we read off
\begin{equation}
f_{kn} = {\cal F}_{kn}(\Omega_{kn})/\Gamma\;.
\label{eq:fourier}
\end{equation}
We use this form of the Fourier expansion coefficients in all of our
calculations.

\section{Examples}
\label{sec:example}

In this section we use the methods described above in two test cases.
We first decompose and reconstruct a simple function of the form
$f(r,\theta)$.  We then show how to decompose a more complicated
function which appears in black hole perturbation calculations.

\subsection{Simple test case}
\label{ssec:reconstruction}

We now test our prescription by computing the expansion coefficients
of a test function and showing that the reconstructed time series
agrees with the original function.  Our test function is $\zeta\equiv
r\cos\theta$, where $r$ and $\theta$ are the Boyer-Lindquist
coordinates.  Our calculations are actually performed using the tips
given in Appendix {\ref{app:practical}}.  In particular, we map the
radial coordinate $r$ to a coordinate $\psi$ defined via
\begin{equation}
r = \frac{pM}{1 + \varepsilon\cos\psi}\;,
\label{eq:r_psi}
\end{equation}
a reparameterization commonly used in Newtonian orbital dynamics
{\cite{marion_thornton}}.  Whereas $r$ oscillates from periapsis
[$r_{\rm peri} = pM/(1 + \varepsilon)$] to apoapsis [$r_{\rm ap} =
pM/(1 - \varepsilon)$] and back, $\psi$ winds secularly from $0$
(periapsis) to $\pi$ (apoapsis) and beyond.  We truncate all infinite
sums at some finite value $N$, discussed below.  We also map the
$\theta$ motion to a coordinate $\chi$ via
\begin{equation}
\cos\theta = \sqrt{z_-(a,E,\iota)}\cos\chi\;,
\end{equation}
where $z_-(a,E,\iota)$ is one root of a quadratic equation defined in
Appendix {\ref{app:practical}}.  The inclination angle $\iota$ relates
the angular momentum $L_z$ to the Carter constant $Q$:
\begin{equation}
\cos\iota = \frac{L_z}{\sqrt{L_z^2 + Q}}\;.
\end{equation}
Although $\iota$ is not quite the geometrical angle describing an
orbit's excursion from the equatorial plane, it is closely related,
and has other convenient properties (see, e.g., Ref.\ {\cite{hb}} for
further discussion).

It's worth noting that the parameterization (\ref{eq:r_psi}) makes
manifestly clear that the radial motion has a slowly converging
Fourier expansion for large eccentricity.  Using the binomial
expansion,
\begin{equation}
r = pM\sum_{n = 0}^\infty (-1)^n \varepsilon^n\cos^n\psi\;.
\label{eq:r_psi_binomial}
\end{equation}
The amplitude of radial harmonic $n$ is roughly $\varepsilon$ smaller
than the amplitude of harmonic $n - 1$.  When we truncate our sums at
some finite value $N$, we expect that a reconstructed time series will
have a fractional error of order $\varepsilon^{N+1}$ (the amplitude of
the next neglected coefficient).  Hence, many harmonics will be needed
as $\varepsilon$ approaches 1.  This slow convergence has been noted
in studies of gravitational radiation reaction on eccentric black hole
orbits {\cite{ckp,gk}}.

\begin{figure}
\epsfig{file = 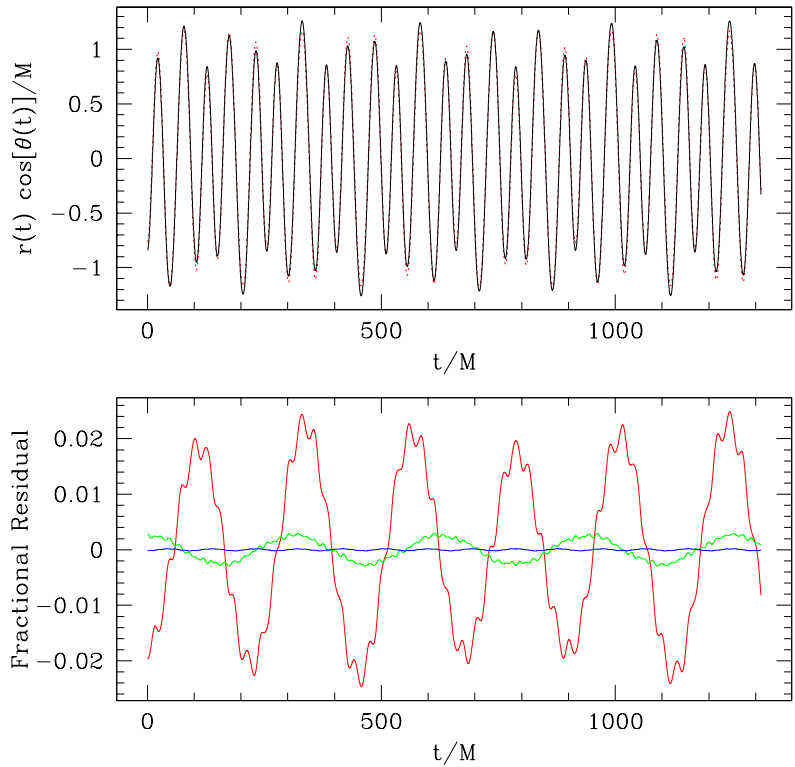, width = 11cm}
\caption{The function $\zeta(t) = r(t)\cos[\theta(t)]$ for orbits
about a black hole with $a = 0.9M$, and with parameters $p = 3$,
$\varepsilon = 0.2$, $\iota = 20^\circ$.  Top panel shows this
function computed directly from the geodesic equations (solid black
line) as well as the time series constructed using Eq.\
(\ref{eq:fourier}) (dotted line; red in color version).  In the
reconstructed time series $\zeta_{\rm rec}(t)$, we have truncated the
infinite sums at $N = 1$.  We find remarkably good agreement despite
the small number of terms kept in the sum.  The bottom panel shows the
fractional residual, $[\zeta(t) - \zeta_{\rm rec}(t)]/\zeta_{\rm
max}$, for several values of $N$.  The largest differences (red in
color version) are for $N = 1$, and have a magnitude of about $0.025$.
The next largest (green in color version) are for $N = 2$ and have a
magnitude of about $0.003$.  The smallest differences (blue in color
version) are for $N = 3$ and have a magnitude of about $0.0002$.}
\label{fig:p3}
\end{figure}

The top panel in Figs.\ {\ref{fig:p3}} and {\ref{fig:p4}} show the
function $\zeta(t)$ computed in two different ways.  The solid black
line shows $\zeta(t)$ constructed by direct integration of the
geodesic equations; the dotted line (red in the color version) shows
$\zeta(t)$ reconstructed from a Fourier expansion using Eq.\
(\ref{eq:fourier}).  The lower panel of these figures shows the
fractional residual, $[\zeta(t) - \zeta_{\rm rec}(t)]/\zeta_{\rm
max}$, where $\zeta_{\rm rec}(t)$ is the reconstructed timeseries and
$\zeta_{\rm max} = r_{\rm max}\cos\theta_{\rm min}$.

\begin{figure}
\epsfig{file = 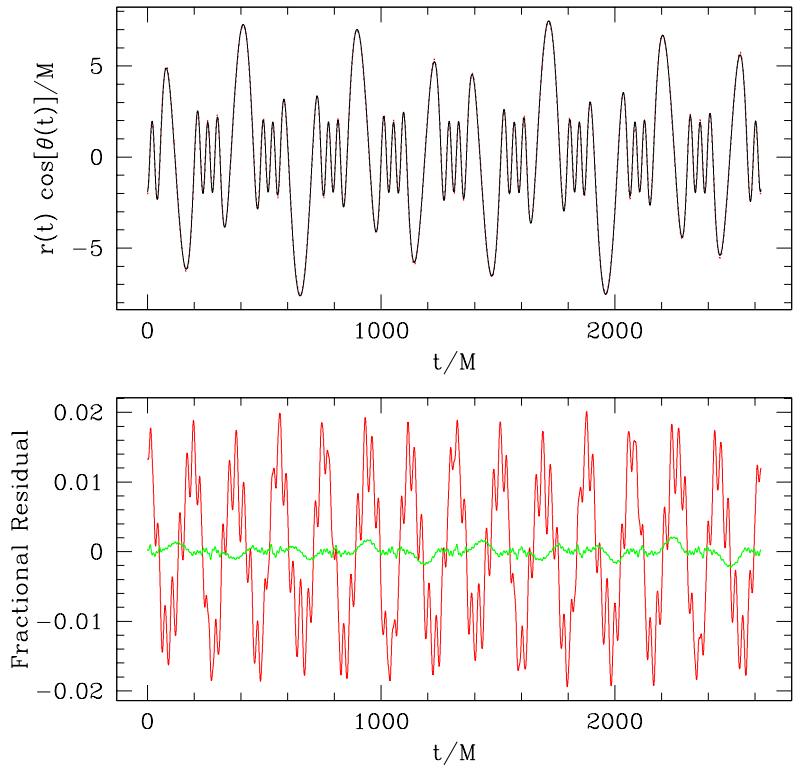, width = 11cm}
\caption{The function $\zeta(t)$ for orbits about a black hole with $a
= 0.9M$ and with parameters $p = 4$, $\varepsilon = 0.6$, $\iota =
50^\circ$.  Top panel shows this function computed directly from the
geodesic equations (solid black line) plus the time series constructed
using Eq.\ (\ref{eq:fourier}) (dotted line; red in color version).  We
have truncated sums in the time series at $N = 4$.  We begin to see
the need to keep a large number of terms at this relatively large
eccentricity.  The bottom panel shows the fractional residual for
several values of $N$.  The largest differences (red in color version)
are for $N = 4$, and have a magnitude of about $0.02$.  The smaller
differences (green in color version) are for $N = 8$ and have a
magnitude of about $0.002$.}
\label{fig:p4}
\end{figure}

Figure {\ref{fig:p3}} compares $\zeta(t)$ and $\zeta_{\rm rec}(t)$ for
an orbit with $p = 3$, $\varepsilon = 0.3$, $\iota = 20^\circ$; the
black hole's spin parameter $a = 0.9M$.  The reconstructed timeseries
converges to $\zeta(t)$ rather quickly: even when the infinite sums
are truncated at $N = 1$, the maximum deviation is only a few percent;
the difference between $\zeta(t)$ and $\zeta_{\rm rec}(t)$ for $N = 1$
is barely discernible on the plot.  The convergence is quite a bit
slower when $\varepsilon = 0.6$.  The motion shown in Fig.\
{\ref{fig:p4}} is for an orbit with $p = 4$, $\varepsilon = 0.6$, and
$\iota = 50^\circ$.  The timeseries is within $2\%$ of $\zeta(t)$ when
$N = 4$; reducing the error by a further factor of ten requires
increasing $N$ to 8.  It's worth noting that, even when the amplitude
error is relatively large, the phase alignment of $\zeta_{\rm rec}(t)$
with $\zeta(t)$ appears to be very good.  This bodes well for problems
that rely on an accurate phase match between data and a model (or
template).

Notice, in both Figs.\ (\ref{fig:p3}) and (\ref{fig:p4}), that the
differences between $\zeta(t)$ and the reconstructed time series are
{\it not} precisely periodic: the wiggles in the lower panels of these
figures do not repeat themselves regularly.  This is a manifestation
of the quasi-periodic nature of the orbital motion.  As more terms are
kept in these sums, we are more successful at capturing this
quasi-periodic motion, and the magnitude of these wiggles quickly
becomes small.

\subsection{Black hole perturbations}
\label{ssec:black hole perturbations}

We now apply these techniques to a problem taken from black hole
perturbation theory {\cite{teuk73}}.  Our goal is to understand how to
decompose a complex function $\psi_4$ which describes how a small body
perturbs the spacetime curvature of a Kerr black hole.  From $\psi_4$,
one can extract information about gravitational-wave emission and
radiative backreaction on small compact objects orbiting massive black
holes --- extreme mass ratio binaries.  When frequency domain
perturbation theory is used to study this problem, $\psi_4$ is
expanded in multipoles and in a harmonic series of the fundamental
orbital frequencies.  The gravitational waves generated by the system
and their backreaction onto the orbit can then be extracted from that
harmonic/multipolar expansion.  Aside from being of great interest to
the current authors, this example nicely illustrates the principles of
this Fourier decomposition for functionals more complicated than the
previous simple example.  We will not dwell too much on the
mathematics of black hole perturbation theory, but will point the
reader to references where appropriate.

The function $\psi_4$ can decomposed into multipoles as
{\cite{teuk73}}
\begin{equation} \label{psi_4}
\psi_4(t_f,r_f,\theta_f,\phi_f) = \rho^{-4} \sum_{lm} \int d\omega\,
R_{lm}(r_f,\omega) S_{lm}(\theta_f,a\omega) e^{i(m\phi_f - \omega t_f)}\;,
\end{equation}
where the angular function $S$ is a spin weighted spheroidal harmonic,
and the radial function $R$ is a solution of a second order ordinary
differential equation known as the Teukolsky equation (see Refs.\
\cite{teuk73,ckp,gk,shibata,msstt,paperI,dh_prep} for a detailed
discussion of the Teukolsky equation).  The subscript $f$ on the
coordinates is a reminder that $(t_f,r_f,\theta_f,\phi_f)$ denotes a
field point.  Coordinates without the subscript will refer to the
location of a body orbiting the black hole.  We will now rewrite
$\psi_4$ completely in terms of sums, eliminating the need for the
integral over $\omega$.

The radial function can be written in the form
\cite{ckp,gk,shibata,msstt,paperI}
\begin{equation}
R_{lm}(r_f,\omega) = Z^{H}_{lm}(r_f,\omega)R^{\infty}_{lm}(r_f,\omega)
+ Z^{\infty}_{lm}(r_f,\omega)R^{H}_{lm}(r_f,\omega)\;,
\end{equation}
where $R^{H,\infty}_{lm}(r_f,\omega)$ are the two independent
solutions to the source-free Teukolsky equation and where the
functions $Z$ are
\begin{equation} \label{short}
Z^\star_{lm}(r_f,\omega) =
\int dt\,e^{i[\omega t - m\phi(t)]} I_{lm}^\star[r(t),\theta(t),r_f,\omega]
\end{equation}
for $\star = H,\infty$.  The function
$I_{lm}^\star[r(t),\theta(t),r_f, \omega]$ depends upon the orbital
worldline of the body perturbing the black hole spacetime.  See Ref.\
{\cite{gk}} for discussion in the case of a body in an equatorial,
eccentric orbit; see Ref.\ {\cite{paperI}} for the case of a body in
an orbit that is inclined but of constant radius.  (The general case,
for orbits that are inclined and eccentric, is in preparation
{\cite{dh_prep}}.)

We next rewrite Eq.\ (\ref{short}) as an integral over $\lambda$
\begin{equation} \label{new short}
Z^\star_{lm}(r_f,\omega) =
\int d\lambda\,e^{i[\omega t(\lambda) - m\phi(\lambda)]} 
{\cal I}_{lm}^\star[r(\lambda),\theta(\lambda),r_f,\omega]\;,
\end{equation}
where ${\cal I}_{lm}^\star = I_{lm}^\star dt/d\lambda$. Now we insert
Eqs.\ (\ref{eq:tlamb}) and (\ref{eq:philamb}) into (\ref{new short})
so that we have
\begin{equation} 
Z^\star_{lm}(r_f,\omega) =
\int d\lambda\ e^{i[\omega \Gamma \lambda - m \Upsilon_\phi \lambda]}
J_{lm}^\star[r(\lambda),\theta(\lambda),r_f,\omega]\;,
\label{eq:Zlm_of_r_omega}
\end{equation}
where 
\begin{equation} \label{new short 2}
J_{lm}^\star[r(\lambda),\theta(\lambda),r_f,\omega] = 
{\cal I}_{lm}^\star[r(\lambda),\theta(\lambda),r_f,\omega]
e^{ i\{\omega \Delta t[r(\lambda),\theta(\lambda)]
- m \Delta\phi[r(\lambda),\theta(\lambda)]\} }\;,
\end{equation} 
with $\Delta t$ and $\Delta\phi$ given by Eqs.\ (\ref{eq:deltatlamb})
and (\ref{eq:deltaphilamb}).  Since $J_{lm}^\star$ depends on
$\lambda$ only through $r(\lambda)$ and $\theta(\lambda)$, it can be
expanded as
\begin{equation} 
J_{lm}^\star[r(\lambda),\theta(\lambda),r_f,\omega] = \sum_{kn}
J_{lmkn}^\star(r_f,\omega) e^{-i\Upsilon_{kn}\lambda}\;.
\end{equation} 
Putting this into Eq.\ (\ref{eq:Zlm_of_r_omega}) and performing the
integral gives
\begin{eqnarray}
Z^\star_{lm}(r_f,\omega) &=& 
2\pi\sum_{jk} J^\star_{lmkn}(r_f,\omega)
\delta(\omega \Gamma - \Upsilon_{mkn})
\nonumber\\
&=& \frac{2\pi}{\Gamma}\sum_{jk} J^\star_{lmkn}(r_f,\omega)
\delta(\omega - \Omega_{mkn})\;,
\label{zdelta}
\end{eqnarray}
where 
\begin{equation} \label{omega_mkn}
\Omega_{mkn} = \Upsilon_{mkn}/\Gamma = m\Omega_\phi + k\Omega_\theta + n\Omega_r\;.
\end{equation}
Finally, when we substitute Eq.\ (\ref{zdelta}) into Eq.\
(\ref{psi_4}) we obtain
\begin{equation} \label{end result}
\psi_4(t_f,r_f,\theta_f,\phi_f) = \frac{1}{\rho^4}
\sum_{lmkn}R_{lmkn}(r_f)S_{lmkn}(\theta_f)e^{i(m\phi_f -
\Omega_{mkn}t_f)}\;,
\end{equation}
where
\begin{eqnarray}
S_{lmkn}(\theta_f) &=& S_{lm}(\theta_f,a\Omega_{mkn})
\nonumber\\
R_{lmkn}(r_f) &=& Z^{H}_{lmkn}(r_f)R^{\infty}_{lm}(r_f,\Omega_{mkn}) +
        Z^{\infty}_{lmkn}(r_f)R^{H}_{lm}(r_f,\Omega_{mkn})
\nonumber\\
Z^\star_{lmkn}(r_f) &=& 2\pi
J^\star_{lmkn}(r_f,\Omega_{mkn})/\Gamma\;.
\end{eqnarray}

\section{Conclusion}
\label{sec:conclude}

With the techniques described in this paper, it should now be a
relatively simple matter to describe functions of Kerr black hole
orbits in the frequency domain.  Although the motion is not truly
periodic with respect to observer time $t$, it {\it is} periodic with
respect to Mino time $\lambda$.  It is thus quite simple to represent
functions using frequencies $\Upsilon$ conjugate to Mino time.  By
using the fact that observer time $t$ is itself periodic with respect
to Mino time (after subtracting the secularly growing contribution),
it is straightforward to convert the $\lambda$-Fourier expansion into
a $t$-Fourier expansion.

As discussed in the Introduction, these techniques could find useful
application to a variety of astrophysical problems involving Kerr
black holes.  One that is of particular interest to us is the problem
of describing gravitational-wave emission from extreme mass ratio
binaries {\cite{ckp,gk,shibata,msstt,paperI}}.  Such systems are
expected to be observable for future space based gravitational wave
detectors.  It should now be fairly straightforward to extend current
black hole perturbation theory codes to handle the very interesting
case of generic orbits --- binaries in which the small body has both
inclination with respect to the equatorial plane and non-zero
eccentricity {\cite{dh_prep}}.  If we had been unable to exploit the
discrete harmonic structure of these systems, such a generalization
would have had an enormous computational cost.  Combining that
analysis with a scheme to compute the evolution of the Carter constant
(using a rigorous computation of a self force {\cite{mino}}, or
perhaps using a cruder approximation {\cite{ghk}}), it should then be
possible to construct, in the adiabatic limit, the inspiral worldlines
and waveforms followed by bodies spiraling into massive black holes.

\begin{acknowledgments}
We are grateful to Marc Favata, \'Eanna Flanagan, and \'Etienne Racine
for valuable discussions that led to this analysis.  We also thank
Wolfram Schmidt for comments on an earlier version of this paper.
This work was supported at Cornell by NSF Grant PHY-0140209 and the
NASA/New York Space Grant Consortium, and at MIT by NASA Grant
NAGW-12906 and NSF Grant PHY-0244424.
\end{acknowledgments}

\appendix

\section{Practical evaluation of the Kerr geodesics}
\label{app:practical}

In this appendix, we present some useful tools for handling functions
of Kerr geodesics.  A difficulty often encountered in evaluating these
orbital motions is due to the presence of turning points in the
motion: as the radial motion approaches periapsis and apoapsis,
$dr/d\mbox{``time''}$ passes through zero and switches sign
(regardless of which time variable one uses).  As the derivative
approaches zero, one typically finds in a numerical evaluation that
small stepsizes are needed to resolve the changing derivative;
precision can be badly degraded in this case.  The $\theta$ behavior
exhibits similar behavior due to the turning points at $\theta_{\rm
min}$ and $\pi - \theta_{\rm min}$.

A simple way to solve this behavior is to work with a functional form
that automatically builds in the correct behavior as the turning
points are approached.  We first describe the transformation used to
describe the $\theta$ motion.  The core idea of this transformation
has been known for quite some time {\cite{wilkins}}, and has been used
extensively in work on circular Kerr black hole orbits
{\cite{paperI}}; it turns out to be particularly simple to use when
studying geodesics parameterized by Mino time.  We then show a simple
transformation that greatly simplifies the description of the radial
motion.  This transformation has also been used quite a bit in
previous work {\cite{ckp,gk}}, but is worth discussing in the context
of the Mino-time parameterization.

\subsection{Motion in $\theta$}
\label{app:thetamotion}

We begin transforming the $\theta$ motion by first defining the
variable $z = \cos^2\theta$.  Equation (\ref{eq:thetalamb}) becomes
\begin{eqnarray}
\frac{d\theta}{d\lambda}
&=& \pm{\sqrt\frac{z^2\left[a^2(1 - E^2)\right] - z\left[Q + L_z^2 +
a^2(1 - E^2)\right] + Q}{1 - z}}
\nonumber\\
&=& \pm{\sqrt\frac{\beta(z_+ - z)(z_- - z)}{1 - z}}\;.
\label{eq:thetalamb2}
\end{eqnarray}
The plus sign corresponds to motion from $\theta_{\rm min}$ to $\pi -
\theta_{\rm min}$, and vice versa for the minus sign.  We have defined
$\beta = a^2(1 - E^2)$; $z_\pm$ are the two roots of the quadratic in
the top line of Eq.\ (\ref{eq:thetalamb2}).

We next define the variable $\chi$: $z = z_-\cos^2\chi$.  As $\chi$
varies from $0$ to $2\pi$, $\theta$ oscillates through its full range
of motion, from $\theta_{\rm min}$ to $\pi - \theta_{\rm min}$ and
back.  Examining $dz/d\theta$ and $dz/d\chi$ we see that
\begin{eqnarray}
\frac{d\chi}{d\theta} &=& \sqrt{\frac{1 - z}{z_- - z}}\;,
\qquad 0 \le \chi \le \pi\;;
\nonumber\\
&=& -\sqrt{\frac{1 - z}{z_- - z}}\;,
\qquad \pi \le \chi \le 2\pi\;.
\label{eq:dchidtheta}
\end{eqnarray}
Combining Eqs.\ (\ref{eq:thetalamb2}) and (\ref{eq:dchidtheta}), we obtain
the geodesic equation for $\chi$:
\begin{eqnarray}
\frac{d\chi}{d\lambda} &=& \sqrt{\beta(z_+ - z)}
\nonumber\\
&=& \sqrt{\beta(z_+ - z_-\cos^2\chi)}\;.
\label{eq:chilamb}
\end{eqnarray}

Using Eq.\ (\ref{eq:chilamb}), it is straightforward to find
$\lambda$ for all $\chi$.  First, define
\begin{equation}
\lambda_0(\chi) = \frac{1}{\sqrt{\beta z_+}}\left[K(\sqrt{z_-/z_+}) -
F(\pi/2 - \chi,\sqrt{z_-/z_+})\right]\;;
\label{eq:lambda0_chi}
\end{equation}
note that
\begin{equation}
\lambda_0(\pi/2) = \frac{1}{\sqrt{\beta z_+}}K(\sqrt{z_-/z_+})\;.
\label{eq:lambda0_chi_comp}
\end{equation}
In these equations, the function $F(\varphi, k)$ is the incomplete
elliptic integral of the first kind, and $K(k)$ is the complete
elliptic integral of the first kind (using the notation of
{\cite{numrec}}).  Then,
\begin{eqnarray}
\lambda(\chi) &=& \lambda_0(\chi)
\qquad 0 \le \chi \le \pi/2
\nonumber\\
&=& \frac{2}{\sqrt{\beta z_+}}K(\sqrt{z_-/z_+}) - \lambda_0(\pi - \chi)
\qquad \pi/2 \le \chi \le \pi
\nonumber\\
&=& \frac{2}{\sqrt{\beta z_+}}K(\sqrt{z_-/z_+}) + \lambda_0(\chi - \pi)
\qquad \pi \le \chi \le 3\pi/2
\nonumber\\
&=& \frac{4}{\sqrt{\beta z_+}}K(\sqrt{z_-/z_+}) - \lambda_0(2\pi - \chi)
\qquad 3\pi/2 \le \chi \le 2\pi\;;
\label{eq:lambda_of_chi}
\end{eqnarray}
also
\begin{equation}
\Lambda_\theta = \frac{4}{\sqrt{\beta z_+}}K(\sqrt{z_-/z_+})\;.
\label{eq:LambdaTheta2}
\end{equation}
This form of $\Lambda_\theta$ is perfectly well behaved even for
orbits that are confined to the equatorial plane ($\theta_{\rm min} =
\pi/2$); this is not the case for the original form
(\ref{eq:LambdaTheta}).

By combining Eqs.\ (\ref{eq:freqdef}), (\ref{eq:chilamb}), and
(\ref{eq:LambdaTheta2}) it is trivial to change variables so that
integrals of $w^\theta$ become integrals over $\chi$:
\begin{eqnarray}
w^\theta(\chi) &=& \Upsilon_\theta\lambda(\chi)\;;
\label{eq:wthetachi}\\
\frac{dw^\theta}{d\chi}
&=& \Upsilon_\theta\frac{d\lambda}{d\chi}
\nonumber\\
&=& \frac{2\pi}{\Lambda_\theta}\frac{1}{\sqrt{\beta(z_+ -
z_-\cos^2\chi)}}
\nonumber\\
&=& \frac{\pi}{2K(\sqrt{z_-/z_+})}\frac{1}{\sqrt{1 -
\left(z_-/z_+\right)\cos^2\chi}}\;.
\label{eq:dwthetadchi}
\end{eqnarray}
Equations (\ref{eq:wthetachi}) and (\ref{eq:dwthetadchi}) are used in
our applications to perform all integrals with respect to the angle
variable $w^\theta$.

\subsection{Motion in $r$}
\label{app:rmotion}

We use a similar trick to simplify the radial motion.  First, we
reparameterize the instantaneous orbital radius as
\begin{equation}
r = \frac{pM}{1 + \varepsilon\cos\psi}\;.
\label{eq:psi_def}
\end{equation}
Such a reparameterization is commonly used to study Keplerian orbits
in Newtonian theory {\cite{marion_thornton}}; though relativistic
orbits are not closed ellipses, the form (\ref{eq:psi_def}) remains
very useful.  The parameter $\varepsilon$ can thus be interpreted as
the eccentricity, $\psi$ as the orbital anomaly, and $p$ as the
semi-latus rectum.  As $\psi$ varies from $0$ to $\pi$, $r$ varies
from periapsis (closest approach) to apoapsis (furthest distance):
\begin{eqnarray}
r_{\rm peri} &=& \frac{pM}{1 + \varepsilon}\;,
\label{eq:r_peri}\\
r_{\rm ap} &=& \frac{pM}{1 - \varepsilon}\;.
\label{eq:r_ap}
\end{eqnarray}

To proceed, we must do some massaging of the function $R(r)$ defined
in Eq.\ (\ref{eq:rdot}).  It is a quartic function of $r$, and thus
has 4 roots:
\begin{eqnarray}
R(r) &=& (E^2 - 1)r^4 + 2M r^3 +[a^2(E^2 - 1) - L_z^2 - Q]r^2
+ 2M[Q + (aE - L_z)^2]r - a^2 Q
\nonumber\\
&=& (1 - E^2)(r_1 - r)(r - r_2)(r - r_3)(r - r_4)\;.
\label{eq:R_roots}
\end{eqnarray}
The second line of Eq.\ (\ref{eq:R_roots}) is written in a way that is
manifestly positive for bound orbits ($E < 1$).  The roots are ordered
such that $r_1 \ge r_2 \ge r_3 \ge r_4$; bound motion occurs for $r_1
\ge r \ge r_2$.  From these definitions, it is clear that $r_1 \equiv
r_{\rm ap}$, and $r_2 \equiv r_{\rm peri}$.

The radii $r_3$ and $r_4$ do not correspond to turning points of the
small body's motion, but of course still represent zeros of the
function $R$.  (In fact, $r_4$ is typically inside the event horizon;
when $Q = 0$ or $a = 0$, $r_4 = 0$.)  It turns out to be useful to
remap these radii as follows:
\begin{eqnarray}
r_3 &=& \frac{p_3 M}{1 - \varepsilon}\;,
\label{eq:p3}\\
r_4 &=& \frac{p_4 M}{1 + \varepsilon}\;.
\label{eq:p4}
\end{eqnarray}
This remapping is simply for mathematical convenience; the parameters
$p_3$ and $p_4$ have no particular physical meaning.

It is now a simple matter to derive the geodesic equation for $\psi$:
\begin{eqnarray}
\frac{d\psi}{d\lambda} &=&
\frac{M\sqrt{1 - E^2}\left[(p - p_3) - \varepsilon(p +
p_3\cos\psi)\right]^{1/2} \left[(p - p_4) + \varepsilon(p -
p_4\cos\psi)\right]^{1/2}}{1 - \varepsilon^2}
\nonumber\\
&\equiv& P(\psi);.
\label{eq:psilamb}
\end{eqnarray}

As with the $\chi$ reparameterization of the $\theta$ motion, it is
straightforward to find $\lambda(\psi)$ using Eq.\ (\ref{eq:psilamb}):
\begin{equation}
\lambda(\psi) = \int_0^\psi \frac{d\psi'}{P(\psi')}\;.
\label{eq:lambda_of_psi}
\end{equation}
In our applications, we evaluate this integral numerically.  It is
possible that an analytic form could be found in terms of elliptic
integrals (though it appears to require more algebraic fortitude than
these authors could muster).  In any practical application, it is
unlikely that such a form will be more useful or accurate than a
numerical evaluation of (\ref{eq:lambda_of_psi}).

Note in particular that
\begin{equation}
\Lambda_r = \int_0^{2\pi} \frac{d\psi'}{P(\psi')}\;.
\label{eq:LambdaR2}
\end{equation}
This form of $\Lambda_r$ is well-behaved in the limit of circular
orbits.

Finally, we use these results to convert integrals over $w^r$ into
integrals over $\psi$: combining Eqs.\ (\ref{eq:freqdef}),
(\ref{eq:psilamb}), and (\ref{eq:LambdaR2}), we have
\begin{eqnarray}
w^r(\psi) &=& \Upsilon_r\lambda(\psi)\;;
\label{eq:wrpsi}\\
\frac{dw^r}{d\psi}
&=& \Upsilon_r\frac{d\lambda}{d\psi}
\nonumber\\
&=& \frac{2\pi}{\Lambda_r}\frac{1}{P(\psi)}\;.
\label{eq:dwrdpsi}
\end{eqnarray}
We use Eqs.\ (\ref{eq:wrpsi}) and (\ref{eq:dwrdpsi}) to perform all
integrals with respect to $w^r$.


\begin{thebibliography}{99}
\bibitem{sums} We could actually write the expansions of $T$ and
  $\Phi$ as a single sums, rather than a pair of sums.  However, the
  pair of sums usefully emphasizes that in practice one might want to
  truncate the $k$ and $n$ sums at different large numbers --- the
  $\theta$ harmonic series usually converges much faster than the $r$
  series.
\end{thebibliography}
\end{document}